\documentclass[a4paper]{article}

\usepackage{INTERSPEECH2020}
\usepackage{url}
\usepackage[colorlinks,linkcolor=black,anchorcolor=black,citecolor=black,urlcolor=black]{hyperref}
\usepackage{multirow}
\usepackage{color}
\usepackage{siunitx}
\usepackage{enumerate}
\title{Reverberation Modeling for Source-Filter-based Neural Vocoder}
\name{Yang Ai$^1$, Xin Wang$^2$, Junichi Yamagishi$^{2,}$$^3$, Zhen-Hua Ling$^1$}
\address{
  $^1$NELSLIP, University of Science and Technology of China, Hefei, P.R.China\\
  $^2$National Institute of Informatics, Japan, $^3$CSTR, University of Edinburgh, UK}
\email{ay8067@mail.ustc.edu.cn, wangxin@nii.ac.jp, jyamagis@nii.ac.jp, zhling@ustc.edu.cn}

\begin{document}

\maketitle
\begin{abstract}
This paper presents a reverberation module for source-filter-based neural vocoders that improves the performance of reverberant effect modeling. This module uses the output waveform of neural vocoders as an input and produces a reverberant waveform by convolving the input with a room impulse response (RIR). We propose two approaches to parameterizing and estimating the RIR. The first approach assumes a global time-invariant (GTI) RIR and directly learns the values of the RIR on a training dataset. The second approach assumes an utterance-level time-variant (UTV) RIR, which is invariant within one utterance but varies across utterances, and uses another neural network to predict the RIR values.
We add the proposed reverberation module to the phase spectrum predictor (PSP) of a HiNet vocoder and jointly train the model.
Experimental results demonstrate that the proposed module was helpful for modeling the reverberation effect and improving the perceived quality of generated reverberant speech. The UTV-RIR was shown to be more robust than the GTI-RIR to
unknown reverberation conditions and achieved a perceptually better reverberation effect.
\end{abstract}

\vspace{-1mm}
\section{Introduction}
\vspace{-1mm}

Recently several neural autoregressive models such as WaveNet \cite{oord2016wavenet}, SampleRNN \cite{mehri2016samplernn}, and WaveRNN \cite{kalchbrenner2018efficient}, have been proposed for raw audio generation. Their variants, such as knowledge-distilling-based models (e.g.,\ parallel WaveNet \cite{oord2017parallel} and ClariNet \cite{ping2018clarinet}) and flow-based models (e.g.,\ WaveGlow \cite{prenger2018waveglow}), were then proposed to further improve the performance and efficiency. These models can be used as \textit{neural vocoders} \cite{tamamori2017speaker,hayashi2017investigation2,adiga2018use,ai2018samplernn,ai2019dnn,lorenzo2018robust} wherein speech waveforms can be reconstructed from acoustic features for various tasks \cite{liu2018wavenet,kobayashi2017statistical,ling2018waveform}. It was confirmed that these neural vocoders outperform vocoders using classical signal processing techniques. However, some limitations still exist --- either a low generation speed, tricky training process, or complicated model structure.

Motivated by the limitations, new types of neural vocoders, such as the glottal neural vocoder \cite{cui2018new,juvela2018speaker} and LPCNet \cite{valin2019lpcnet}, have been further proposed by combining speech production mechanisms with neural networks, and their performance is impressive. However, all of the above models operate under the autoregressive assumption and are slow in either waveform generation or training. Previously, we proposed non-autoregressive neural source-filter (NSF) \cite{wangNSFall} and HiNet vocoders \cite{ai2020neural,ai2020knowledge}. The NSF vocoder uses dilated convolutions to transform a sine-based source signal into an output waveform, following the idea of the source-filter speech production model \cite{gunnar1960acoustic}.
The HiNet vocoder is composed of an amplitude spectrum predictor (ASP) and a phase spectrum predictor (PSP), where the PSP is built by using the NSF vocoder for better phase recovery. The outputs of the ASP and PSP are combined to recover speech waveforms via short-time Fourier synthesis (STFS). Experimental results show that the NSF and HiNet vocoders can generate waveforms with high quality and high efficiency for speech \cite{ai2020neural,ai2020knowledge} and musical instrument sounds \cite{zhao2019transferring} recorded in acoustically isolated studios.

However, unlike the ideal data for speech or music synthesis, audio signals captured for real-life applications typically contain room reverberation.
The reverberation poses a challenge to non-autoregressive neural vocoders, and the quality of synthesized speech usually degrades.
Recently, Engel \MakeLowercase{\textit{et al.}} tried to introduce a reverberation module with a trainable room impulse response (RIR) into a sinusoidal vocoder \cite{engel2020ddsp}. Their model successfully learned room reverberation effects on a solo violin dataset under a signal reverberation condition. However, learning the reverberation effects in multiple acoustic environments and applying the model for unseen acoustic environments have not yet been investigated, and neither has the model been evaluated on a reverberant speech dataset.

As an initial step towards robust reverberation modeling for speech data, this paper proposes a trainable reverberation module for neural vocoders. This module uses the output waveform of neural vocoders as an input and outputs a reverberant waveform by convolving the input with a RIR. We design two types of neural RIR estimators. One estimates the global time-invariant (GTI) RIR, which is invariant among a whole dataset and is regarded as a trainable variable of a model. This is similar to \cite{engel2020ddsp}. The other infers an utterance-level time-variant (UTV) RIR, which is invariant inside one utterance but varies among different utterances. The UTV-RIRs are predicted by an additional trainable neural network that uses the same conditional features as neural vocoders.
%
%
We add the proposed reverberation module to the PSP of the HiNet vocoder, and experiments are conducted on a multi-speaker reverberant speech database with various types of reverberation conditions, including unseen ones.
Furthermore, a multi-task training strategy that uses both reverberant and corresponding dry waveforms is also investigated.

This paper is organized as follows: In Section \ref{sec: Previous works}, we briefly review the NSF and HiNet vocoders. In Section \ref{sec: Proposed methods}, we give details on our proposed reverberation module and neural RIR estimators. Section \ref{sec: Experiments} reports our experimental results. Conclusions are given in Section \ref{sec: Conclusion}.

\vspace{-1mm}
\section{Brief view of NSF and HiNet vocoders}
\label{sec: Previous works}
\vspace{-1mm}


The NSF models \cite{wangNSFall} generate speech waveforms  from input acoustic features through time-domain non-linear transformations. They include three modules: a conditional module that upsamples input acoustic features such as F0 and mel-spectrogram, a source module that outputs a sine-based source signal given the F0, and a dilated-convolution-based filter module that transforms the source signal into an output waveform. The NSF models are suitable for applications where the users want to precisely control the F0 of the output waveform.


HiNet \cite{ai2020neural} is a neural vocoder that produces speech waveforms from input acoustic features by predicting amplitude and phase spectra hierarchically. The HiNet vocoder consists of two predictors, an ASP and a PSP. The ASP uses acoustic features as input and predicts frame-level log amplitude spectra (LAS). Then, the F0 and LAS predicted by the ASP are sent into the PSP for phase spectra prediction. Finally, the predicted amplitude and phase spectra are combined to reconstruct speech waveforms by STFS.


In our latest work \cite{ai2020knowledge}, the ASP consisted of multiple convolutional layers for converting acoustic features into the LAS. Generative adversarial networks (GANs) were also newly introduced into the ASP; the ASP was used as a generator, and two discriminators were adopted. Both discriminators consisted of multiple convolutional layers, which operate convolution along with either the frequency or time axis of the input LAS, respectively. The training of the ASP is based on a Wasserstein GAN \cite{gulrajani2017improved} loss together with the mean square error (MSE) between the predicted LAS and natural ones.

The PSP conducts two steps: neural waveform generation and phase spectrum extraction. The neural waveform generator was based on a customized NSF vocoder \cite{wangNSFall} with three modifications for better phase recovery: 1) the use of LAS as input, 2) pre-calculation of the initial phase of the sine-based excitation signal for each voiced segment at the training stage, and 3) the use of a combined loss function including MSE on amplitude spectra, waveform loss, and correlation loss.

\vspace{-1mm}
\section{Proposed methods}
\label{sec: Proposed methods}
\vspace{-1mm}

\begin{figure}[t]
    \centering
    \includegraphics[width=1\columnwidth]{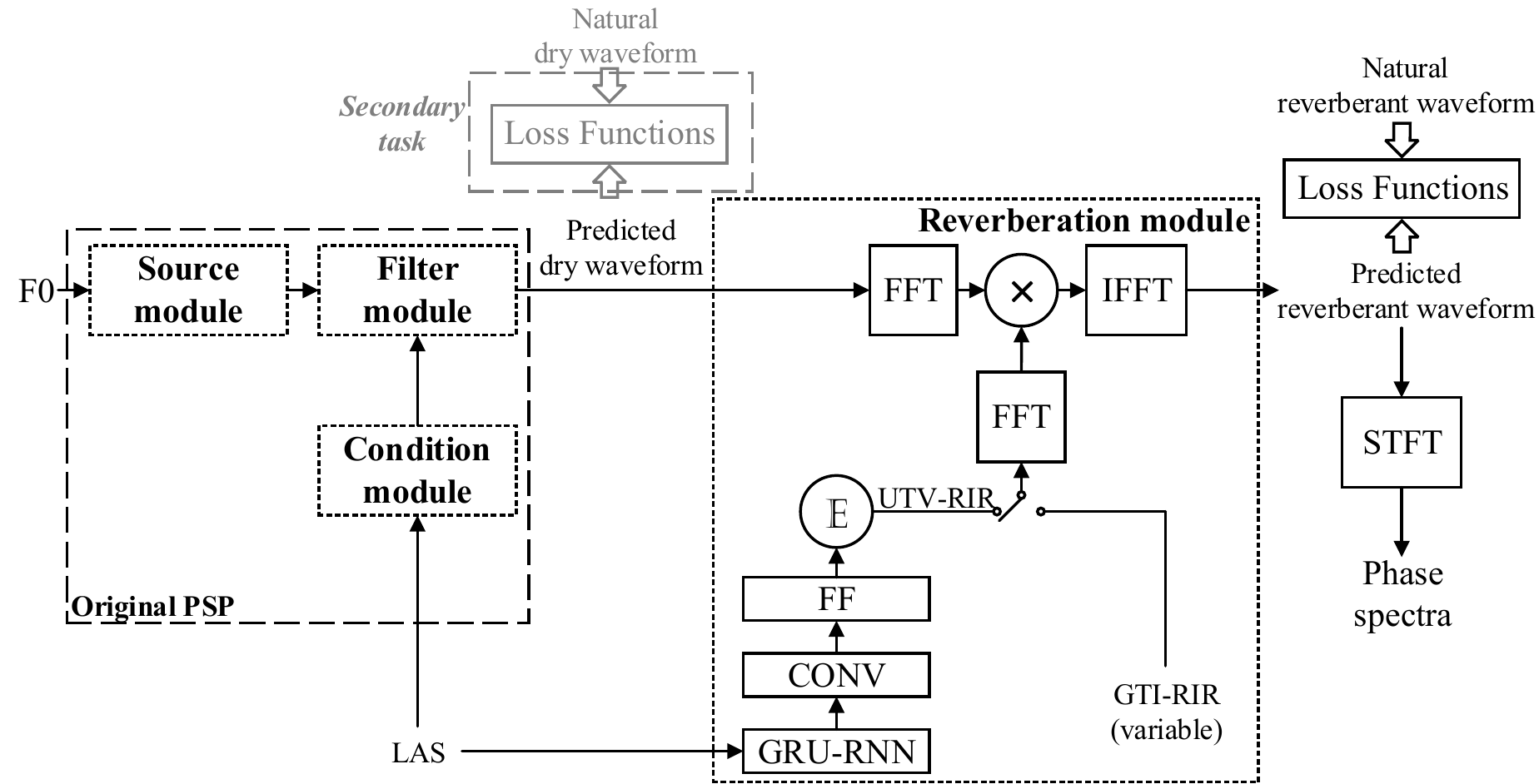}
    \caption{Flowchart of PSP with reverberation module. Here {}{FF}, {}{CONV}, and {}{GRU-RNN} represent feed-forward, convolutional, and unidirectional GRU-based recurrent layers, respectively, and $\times$ and $\mathbb{E}$ denote element-wise product and temporal averaging operation, respectively.}
    \label{fig: PSP}
    \vspace{-5mm}
\end{figure}

When a speech waveform signal $\bm{d}=[d_1,\dots,d_{T}]^\top$ of length $T$ is produced in a closed room, it propagates to an observation point through a direct path, reflects off walls and surrounding objects and becomes a reverberant signal.
By assuming that the RIR of a room can be approximated by the finite impulse response sequence $\bm{h}=[h_1,\dots,h_{L}]^\top$ \cite{naylor2010speech, benesty2007springer}, where $h_1=1$ denotes the direct path, the received reverberant signal  $\bm{r}=[r_1,\dots,r_{T}]^\top$ can be written as
\begin{equation}
\label{equ: reverb}
\bm{r}=\bm{d}* \bm{h}.
\end{equation}
On the basis of this principle, we propose a reverberation module for the HiNet vocoder. This module accepts the output waveform of the PSP in the HiNet vocoder as input, which is illustrated in Fig. \ref{fig: PSP}.\footnote{We also tried to add a reverberation module based on a causal convolution network for the ASP, but it was not effective.}
Although we can directly compute Eq.~(\ref{equ: reverb}) through convolution in the time domain, in order to reduce the computational cost, we implement the convolution in the frequency domain as
\begin{equation}
\label{equ: reverb_frequency}
\bm{r}=\mathcal{F}^{-1}[\mathcal{F}(\bm{d})\odot \mathcal{F}(\bm{h})],
\end{equation}
where $\mathcal{F}$, $\mathcal{F}^{-1}$, and $\odot$ represent the FFT, inverse FFT, and element-wise product, respectively.

There are two ways to parameterize and estimate the value of the RIR $\bm{h}$\footnote{We also tried to parameterize the RIR as an exponentially decaying function with a trainable decay rate, but the learned RIR was intractable.}:
\begin{itemize}
\setlength{\leftskip}{-0.5cm}
\item \textbf{Global time-invariant (GTI) RIR}: inspired by DDSP \cite{engel2020ddsp}, the RIR $\bm{h}$ is assumed to be time-invariant and shared for all speech data, and the values of its coefficients $\{h_1,\dots,h_{L}\}$ are learned from the training data. Note that, because $h_1=1$, only $L-1$ elements in $\bm{h}$ need to be learned.
\item \textbf{Utterance-level time-variant (UTV) RIR}: the RIR $\bm{h}$ is assumed to be invariant for one utterance, but different utterances acquire different $\bm{h}$. The value of $\bm{h}$ is predicted from the input LAS by a small conditional network that consists of trainable recurrent layers, convolutional layers, feed-forward layers, and a temporal average pooling layer as shown in Fig.\ \ref{fig: PSP}. The temporal average pooling layer averages the hidden features of all of the frames and gives a single vector as the predicted $\bm{h}$.
\end{itemize}
The GTI-RIR is expected to be suitable for scenarios where we want to learn the RIR of one time-invariant acoustic environment, while the UTV-RIR is suitable for more general cases where the speech data is recorded in several different acoustic environments.
During training, the reverberation module and the PSP are jointly optimized by a loss function consisting of multi-resolution spectral distortions \cite{wangNSFall} between the output of the reverberation module and the natural reverberant waveform.

For cases where the dry waveforms of the reverberant data are also available (e.g., when reverberation data are generated from clean data through simulation or replaying), we further investigate a multi-task training strategy that uses not only reverberant data but also dry waveforms.
As the gray region in Fig.\ \ref{fig: PSP} shows, the loss function of the secondary task is a combination of MSE on LAS, waveform loss, and correlation loss  \cite{ai2020neural}  between a generated dry waveform and the natural dry waveform. The whole loss function is the sum of the loss functions of the main and secondary tasks.




\vspace{-1mm}
\section{Experiments}
\label{sec: Experiments}
\vspace{-1mm}

\subsection{Data and feature configuration}
\label{subsec: Data and feature configuration}
\vspace{-1mm}

A multi-speaker reverberant speech database\footnote{\url{https://doi.org/10.7488/ds/1425}} \cite{valentini2018speech} was used in our experiments.
From the database, we used a reverberant subset of 28 speakers that contained 11,572 utterances and 18 reverberation types (9 rooms $\times$ 2 microphones positions). We randomly divided this subset into a training set (11,012 utterances) and validation set (560 utterances). Regarding the test set, there were three scenarios below in our experiments:
\begin{enumerate}[\textbf{T}1]
\vspace{-1.2mm}
\item Two unseen speakers' reverberant data with 6 unseen reverberation types (3 rooms $\times$ 2 microphone positions), 824 utterances in total;
\vspace{-1.2mm}
\item Two unseen speakers' reverberant data with the  same 18 reverberation types as in the training set, 832 utterances in total;
\vspace{-1.2mm}
\item Dry speech version of \textbf{T}1.
\end{enumerate}
\vspace{-1mm}

The original 48-kHz waveforms were down sampled to 24-kHz for the experiments.
The acoustic features included the 80-dimensional mel-spectrogram, F0 extracted using YAPPT \cite{kasi2002yet}, and a voiced/unvoiced flag.
The LAS used by HiNet was computed using 2048 FFT points. All features were extracted with a frame shift and length of 12 and 50 ms, respectively.

\vspace{-1mm}
\subsection{Experimental models}
\label{subsec: Experimental models}
\vspace{-1mm}

We compared the following variants in the experiments\footnote{Examples of generated speech can be found at \url{http://home.ustc.edu.cn/~ay8067/reverb/demo.html}. Scripts and toolkits for the NSF model can be found at \url{https://github.com/nii-yamagishilab/project-CURRENNT-scripts}}:

\vspace{0.5mm}
\noindent
{}{\textbf{N-BL}}: The harmonic-plus-noise NSF model \cite{wang2020using} was included as a reference model without the reverberation module. The number of model parameters was around \num{1.2e6}.

\vspace{0.5mm}
\noindent
{}{\textbf{H-BL}}: Baseline HiNet vocoder without the proposed reverberation module.
We used the baseline ASP configuration in our previous work \cite{ai2020knowledge} and added two convolution layers having 512 and 1024 channels, respectively, to the discriminator that operated along with the frequency axis. This was motivated by the increased number of FFT points of the LAS. We adopted the same PSP used as the baseline in our previous work \cite{ai2020knowledge}, but GANs were not used here. The loss function was a combination of MSE on LAS, waveform loss, and correlation loss.
Note that the NSF module in the PSP is slightly different from  {}{\textbf{N-BL}} (see details in \cite{ai2020neural}).
The number of model parameters was around \num{6.2e7} for the ASP and \num{7.2e6} for the PSP.

\vspace{0.5mm}
\noindent
{}{\textbf{H-GTI:}} HiNet with the GTI-RIR-based reverberation module integrated into PSP. The RIR length was 6,000. The model configuration and size were the same as {}{\textbf{H-BL}} except for the increased 5,999 trainable parameters for GIT-RIR.

\vspace{0.5mm}
\noindent
{}{\textbf{H-UTV:}} HiNet with the UTV-RIR-based reverberation module integrated into the PSP. The RIR length was 6,000. The trainable neural network that converts LAS to RIR consisted of a unidirectional GRU layer with 1,024 nodes, a convolution layer with 1,024 nodes and a kernel-size of 11, and a feed-forward linear layer with 5,999 output nodes. Other settings were the same as those of {}{\textbf{H-BL}}.
The number of model parameters for the PSP was increased by \num{2.4e7} compared with {}{\textbf{H-BL}}. Since the UTV-RIR is non-autoregressive, the increased model size did not cause an obvious degradation of generation efficiency.

\vspace{0.5mm}
\noindent
{}{\textbf{H-UTV-MT:}} same as {}{\textbf{H-UTV}} but with the secondary task using dry waveforms during training.

\vspace{-1mm}
\subsection{Main experiments}
\label{subsec: Main experimental results}
\vspace{-1mm}

Our main experiments focused on the reverberation effect and speech quality.
We compared  {}{\textbf{N-BL}},  {}{\textbf{H-BL}},  {}{\textbf{H-GTI}}, and  {}{\textbf{H-UTV}} under testing scenarios \textbf{T}1 and \textbf{T}2 using both objective and subjective evaluations.

\vspace{-1mm}
\subsubsection{Objective evaluation -- T60 comparisons --}
\label{subsubsec: Objective evaluation}
\vspace{-1mm}
\begin{figure}[t]
    \centering
    \includegraphics[width=1.0\columnwidth]{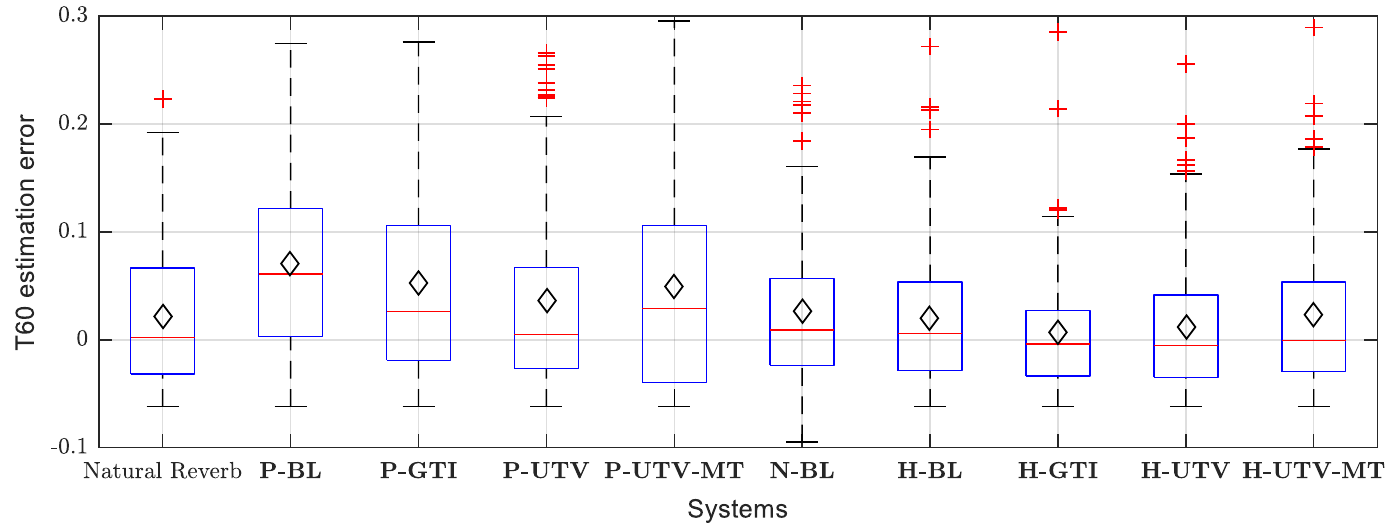}
    \vspace{-5mm}
    \caption{Box plots of T60 estimation errors for utterances with T60n = 0.362s under test scenario \textbf{T}1.}
    \label{fig: T60}
    \vspace{-5mm}
\end{figure}

T60 estimation errors \cite{gaubitch2012performance} were used as the objective metric to evaluate the reverberation effects. T60 is also called the reverberation time, and it is defined as the time it takes for sound to decay by 60 dB after the source has been switched off. We used an open source toolkit \cite{marco2020blind} to blindly estimate T60 from the reverberant speech. The T60 estimation errors were calculated as the difference between the estimated T60 and the ground-truth T60 (T60n) reported in the database paper \cite{valentini2018speech}.
%

We calculated the T60 estimation errors for the output waveforms from all of the experimental models. For reference, we also calculated the errors for the natural reverberant waveform and the output waveform from the PSP in the HiNet models (denoted by {}{\textbf{P-*}}).
Figure \ref{fig: T60} shows box plots of T60 estimation errors for utterances with T60n = 0.362s
under test scenario \textbf{T}1.
Note that the T60 estimated errors for natural reverberant speech were non-zero because blind estimation of T60 is not error-free.

Figure \ref{fig: T60} demonstrates that both {}{\textbf{P-GTI}} and {}{\textbf{P-UTV}} had smaller errors than {}{\textbf{P-BL}}, which indicates the usefulness of the proposed reverberation module in the PSP component of HiNet. Furthermore, {}{\textbf{P-UTV}} had a smaller error than {}{\textbf{P-GTI}}, suggesting  that UTV-RIR is more effective than GTI-RIR in modeling unseen reverberation types.
By comparing {}{\textbf{P-*}} with {}{\textbf{H-*}}, we see that {}{\textbf{H-*}} had smaller errors than {}{\textbf{P-*}}. This suggests that the ASP is able to produce the reverberation effect by a moderate amount even though the ASP has no explicit reverberation module. The performance differences among {}{\textbf{H-*}} vocoders are small.
Additionally we can observe that {}{\textbf{N-BL}} had smaller errors than {}{\textbf{P-BL}} while {}{\textbf{H-BL}} had marginally smaller errors than {}{\textbf{N-BL}}.

\vspace{-1mm}
\subsubsection{Subjective evaluation}
\vspace{-1mm}
We conducted two types of listening tests on the crowdsourcing platform Amazon Mechanical Turk\footnote{\url{https://www.mturk.com}.} with anti-cheating considerations \cite{buchholz2011crowdsourcing} to evaluate the reverberation effect and speech quality, respectively.
In each test, 20 test-set utterances were generated for each test scenario by each experimental model, and these utterances were evaluated by about 40 English native listeners.

\vspace{1mm}
\noindent
\textbf{Reverberation effect:} The first test was a similarity test on the reverberation effect.
Listeners were asked to first listen to the natural dry and reverberant audio tracks. They were then asked to listen to a few test audio tracks and assign a score from 1 to 9 to each, where a higher score denoted a reverberation effect more similar to that in the natural reverberant audio tracks.
The audio tracks generated from the PSP in the HiNet models were directly used for the listening test, and they are denoted as {}{\textbf{P-*}}. Furthermore, to investigate the impact of the proposed reverberation module, the listening test used the audio tracks generated from the PSP after the trained reverberation module was removed, and they are denoted as {}{\textbf{P-*(dry)}}.

\begin{figure}[t]
    \centering
    \includegraphics[width=1.0\columnwidth]{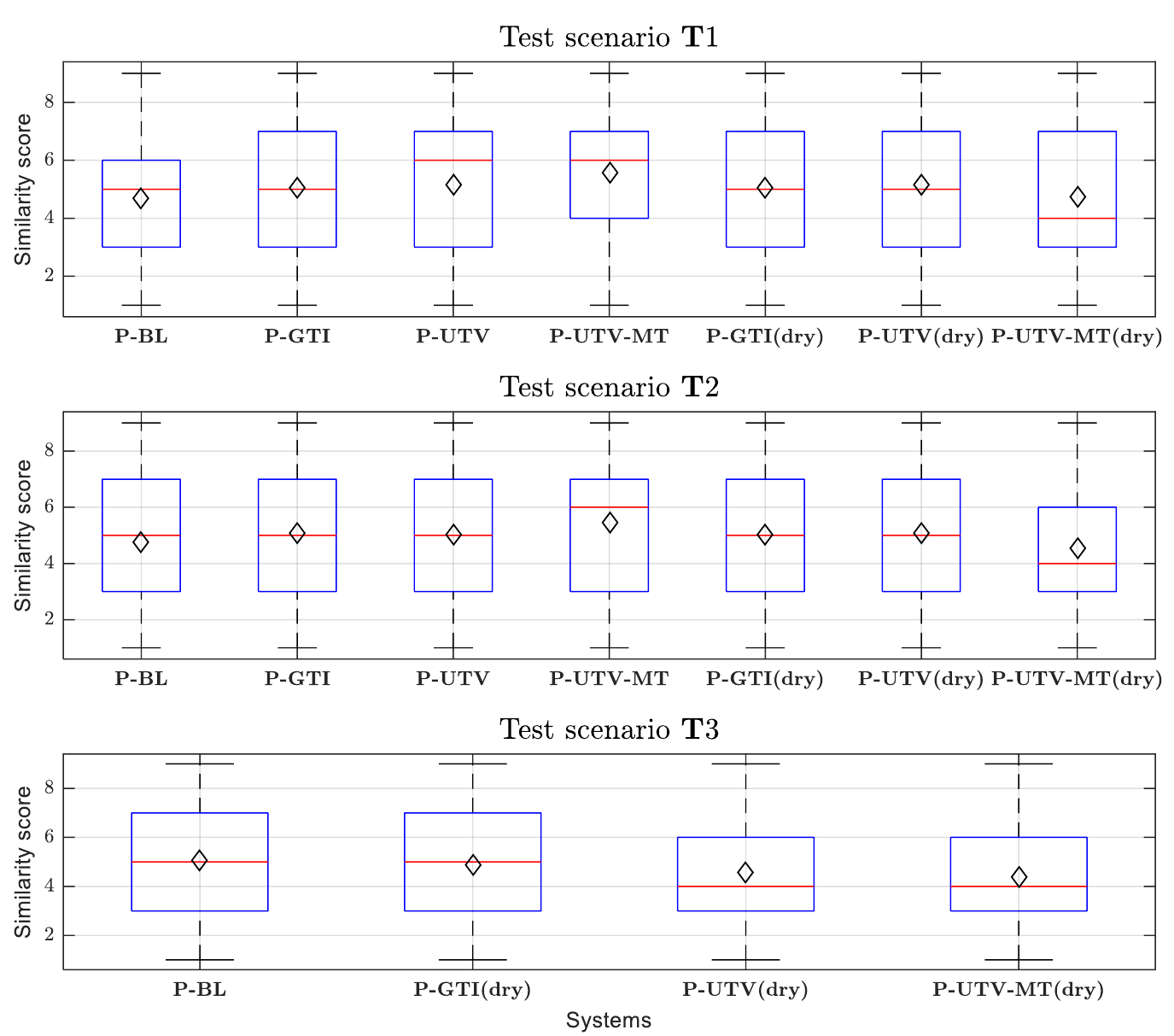}
    \vspace{-5mm}
    \caption{Box plot of reverberant effect similarity scores for all test scenarios. Here, black diamonds and red lines represent mean and median.}
    \label{fig: Reverb_Similarity}
    \vspace{-5mm}
\end{figure}

The results for test scenarios \textbf{T}1 (unseen reverberant type) and \textbf{T}2 (seen reverberant type) are plotted in Figure \ref{fig: Reverb_Similarity}.
As expected, the similarity scores of {}{\textbf{P-GTI}} and {}{\textbf{P-UTV}} had higher means and medians than those of {}{\textbf{P-BL}} in both \textbf{T}1 and \textbf{T}2. This means that the proposed module generated reverberation that was perceptually more similar to the natural reverberant speech. Furthermore, {}{\textbf{P-UTV}} outperformed {}{\textbf{P-GTI}} in \textbf{T}1. These results were consistent with the results for the T60 estimation errors in Section \ref{subsubsec: Objective evaluation}.
For \textbf{T}2, however, {}{\textbf{P-UTV}} did not outperform {}{\textbf{P-GTI}}, which indicates that {}{\textbf{P-UTV}} may be more suitable for unknown reverberation conditions.
Unfortunately, the similarity scores of {}{\textbf{P-*(dry)}} remained similar to {}{\textbf{P-BL}}. It seems that the evaluated models did not have de-reverberation ability, i.e., they could not generate perfect dry waveforms giving reverberant acoustic features. One possible reason may be that the reverberation module was jointly trained with the rest of the network. This point is further investigated together with multi-task learning in the next section.

\begin{figure}[t]
    \centering
    \includegraphics[width=0.8\columnwidth]{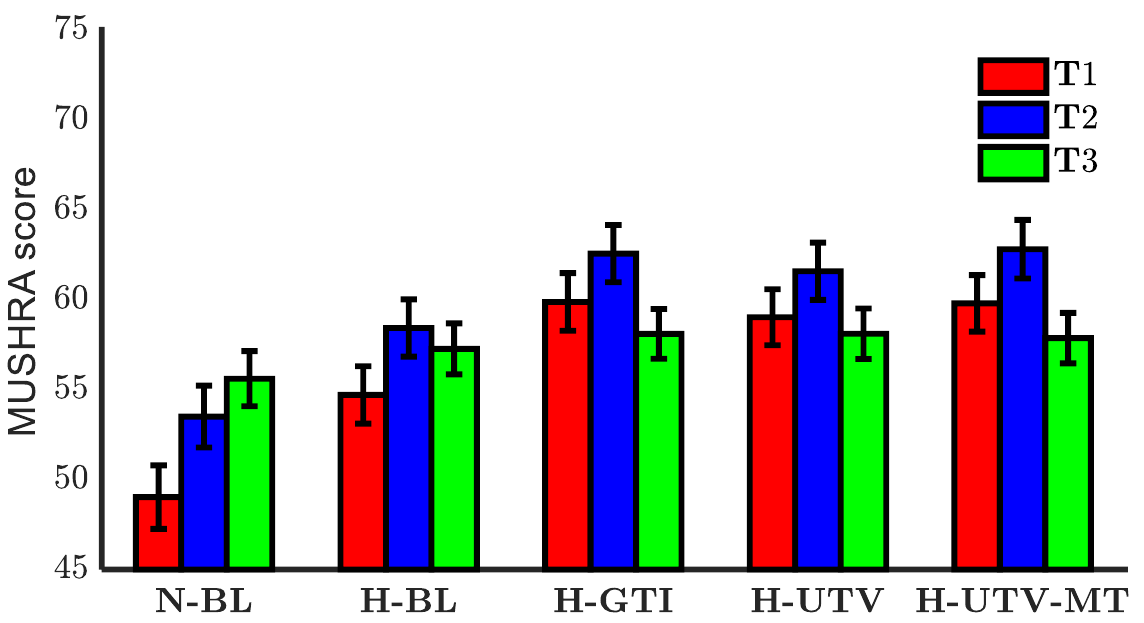}
    \vspace{-3mm}
    \caption{Average MUSHRA scores with 95\% confidence interval for all test scenarios.}
    \label{fig: MUSHRA}
    \vspace{-5mm}
\end{figure}

\vspace{1mm}
\noindent
\textbf{Quality:} The second test was a MUSHRA (MUltiple Stimuli with Hidden Reference and Anchor) test \cite{recommendation2001method} done to compare the quality of the generated waveforms. The average MUSHRA scores and their 95\% confidence intervals are shown in Figure \ref{fig: MUSHRA}.
The reference audio tracks in the MUSHRA test for \textbf{T}1 and \textbf{T}2 were the natural reverberant waveforms.

As Figure \ref{fig: MUSHRA} shows, {}{\textbf{H-GTI}} and {}{\textbf{H-UTV}} had higher MUSHRA scores than {}{\textbf{H-BL}} for both \textbf{T}1 and \textbf{T}2, suggesting that the reverberation module in the PSP was helpful for improving the quality of synthetic speech for HiNet. The MUSHRA scores for \textbf{T}2 were higher than those for \textbf{T}1, which is reasonable since unseen reverberation conditions are more challenging to model. We can also see that the difference between {}{\textbf{H-GTI}} and {}{\textbf{H-UTV}} was not significant. Utterance-dependent RIR estimation seems to be important for modeling multiple reverberation types as the T60 comparison and the similarity test suggest, but it does not improve the perceived quality of generated waveforms.
%
%
Finally, {}{\textbf{H-BL}} outperformed {}{\textbf{N-BL}}, and this indicates that the reverberant speech generated from the HiNet vocoder sounded better than that from the NSF vocoder with the current configurations.

%

\vspace{-1mm}
\subsection{Additional analysis}
\label{subsec: Additional experimental results}
\vspace{-1mm}

Finally, we analyzed two additional configurations.

\vspace{1mm}
\noindent
\textbf{Multi-task training using dry waveforms:} Models using the multi-task training are denoted as {}{\textbf{*-UTV-MT}}. By comparing {}{\textbf{P-UTV-MT}} with {}{\textbf{P-UTV}} in Figure \ref{fig: T60}, we see that using the multi-task training did not reduce the T60 estimation error. However, as the similarity test results in Figure \ref{fig: Reverb_Similarity} show, {}{\textbf{P-UTV-MT}} had a higher mean and median than {}{\textbf{P-UTV}}. Furthermore,
the median of {}{\textbf{P-UTV-MT(dry)}} was 4.0, while that of {}{\textbf{P-UTV-MT}} was 6.0 for \textbf{T}1 and \textbf{T}2, and the differences were larger than those between {}{\textbf{P-UTV}} and {}{\textbf{P-UTV(dry)}}. These results suggest that multi-task training using dry waveforms as additional supervision makes the functional role of the proposed reverberation module more explicit.
Regarding the quality, there was no obvious difference between between {}{\textbf{H-UTV-MT}} and {}{\textbf{H-UTV}} as shown in Figure \ref{fig: MUSHRA}.

\vspace{0.5mm}
\noindent
\textbf{Use of dry acoustic features (\textbf{T}3):} If the proposed framework is well generalized, it should be able to generate dry speech with high quality when we input dry acoustic features. This was the purpose of \textbf{T}3.
The results in Figure \ref{fig: Reverb_Similarity} show that the medians or the mean similarity scores of \textbf{P-GTI(dry)}, \textbf{P-UTV(dry)}, and \textbf{P-UTV-MT(dry)} were lower than those of the corresponding models in \textbf{T}1 and \textbf{T}2. In other words, these models generated waveforms that were perceptually closer to the natural dry waveforms when using dry input acoustic features. These results are encouraging. However, the generated waveforms were not sufficiently close to the natural dry waveforms, so there is still room for improvement.
The results of the quality comparisons are shown in Figure \ref{fig: MUSHRA}. The reference tracks used for the MUSHRA test were natural waveforms without reverberation.
From the MUSHRA listening test, we can see that the quality scores for \textbf{T}3 were similar to those of \textbf{T}1 for unseen conditions.


\vspace{-1mm}
\section{Conclusion}
\label{sec: Conclusion}
\vspace{-1mm}

In this paper, we proposed a neural reverberation module and integrated it into non-autoregressive source-filter-based neural vocoders. The reverberation module uses RIRs for convolving waveforms generated by the vocoders as the standard signal processing method does, but the RIRs are estimated jointly with other parameters of the neural vocoder or predicted by another trainable network. The former approach, called GIT-RIR, uses a globally invariant vector and is directly trained on a reverberant dataset. The latter approach, called UTV-RIR, uses another network to estimate a different RIR for each utterance. We conducted experiments by adding the proposed reverberation module to the PSP of the HiNet vocoder. Objective and subjective evaluation results indicated that the proposed reverberation module is helpful for modeling the reverberation effect and improving the quality of reverberant speech generated by the HiNet vocoder. We also confirmed that the UTV-RIR was better than the GTI-RIR when modeling multiple unseen reverberation types. For future work, we plan to apply the reverberation module to other neural vocoders.


\vspace{1mm}
\footnotesize{
\noindent\textbf{Acknowledgments:}
%
This work was partially supported by a JST CREST Grant (JPMJCR18A6, VoicePersonae project), Japan, MEXT KAKENHI Grants (19K24371, 16H06302, 17H04687, 18H04120, 18H04112, 18KT0051), Japan, and the National Natural Science Foundation of China under Grant 61871358 and the China Scholarship Council (CSC). The experiments were partially conducted on TSUBAME 3.0 of Tokyo Institute of Technology.
}

\newpage

\bibliographystyle{IEEEtran}

\bibliography{mybib}

\begin{thebibliography}{10}
\providecommand{\url}[1]{#1}
\csname url@samestyle\endcsname
\providecommand{\newblock}{\relax}
\providecommand{\bibinfo}[2]{#2}
\providecommand{\BIBentrySTDinterwordspacing}{\spaceskip=0pt\relax}
\providecommand{\BIBentryALTinterwordstretchfactor}{4}
\providecommand{\BIBentryALTinterwordspacing}{\spaceskip=\fontdimen2\font plus
\BIBentryALTinterwordstretchfactor\fontdimen3\font minus
  \fontdimen4\font\relax}
\providecommand{\BIBforeignlanguage}[2]{{%
\expandafter\ifx\csname l@#1\endcsname\relax
\typeout{** WARNING: IEEEtran.bst: No hyphenation pattern has been}%
\typeout{** loaded for the language `#1'. Using the pattern for}%
\typeout{** the default language instead.}%
\else
\language=\csname l@#1\endcsname
\fi
#2}}
\providecommand{\BIBdecl}{\relax}
\BIBdecl

\bibitem{oord2016wavenet}
A.~van~den Oord, S.~Dieleman, H.~Zen, K.~Simonyan, O.~Vinyals, A.~Graves,
  N.~Kalchbrenner, A.~Senior, and K.~Kavukcuoglu, ``{WaveNet: A generative
  model for raw audio},'' \emph{arXiv preprint arXiv:1609.03499}, 2016.

\bibitem{mehri2016samplernn}
S.~Mehri, K.~Kumar, I.~Gulrajani, R.~Kumar, S.~Jain, J.~Sotelo, A.~Courville,
  and Y.~Bengio, ``{{SampleRNN}: {An} unconditional end-to-end neural audio
  generation model},'' in \emph{arXiv preprint arXiv:1612.07837}, 2017.

\bibitem{kalchbrenner2018efficient}
N.~Kalchbrenner, E.~Elsen, K.~Simonyan, S.~Noury, N.~Casagrande, E.~Lockhart,
  F.~Stimberg, A.~Oord, S.~Dieleman, and K.~Kavukcuoglu, ``Efficient neural
  audio synthesis,'' in \emph{Proc. ICML}, 2018, pp. 2410--2419.

\bibitem{oord2017parallel}
A.~v.~d. Oord, Y.~Li, I.~Babuschkin, K.~Simonyan, O.~Vinyals, K.~Kavukcuoglu,
  G.~v.~d. Driessche, E.~Lockhart, L.~C. Cobo, F.~Stimberg \emph{et~al.},
  ``Parallel {W}ave{N}et: Fast high-fidelity speech synthesis,'' in \emph{Proc.
  ICML}, 2018, pp. 3918--3926.

\bibitem{ping2018clarinet}
W.~Ping, K.~Peng, and J.~Chen, ``{ClariNet: Parallel wave generation in
  end-to-end text-to-speech},'' in \emph{arXiv preprint arXiv:1807.07281},
  2019.

\bibitem{prenger2018waveglow}
R.~Prenger, R.~Valle, and B.~Catanzaro, ``{WaveGlow: A Flow-based Generative
  Network for Speech Synthesis},'' in \emph{Proc. ICASSP}, 2019, pp.
  3617--3621.

\bibitem{tamamori2017speaker}
A.~Tamamori, T.~Hayashi, K.~Kobayashi, K.~Takeda, and T.~Toda,
  ``Speaker-dependent {W}ave{N}et vocoder,'' in \emph{Proc. Interspeech}, 2017,
  pp. 1118--1122.

\bibitem{hayashi2017investigation2}
T.~Hayashi, A.~Tamamori, K.~Kobayashi, K.~Takeda, and T.~Toda, ``An
  investigation of multi-speaker training for {W}ave{N}et vocoder,'' in
  \emph{Proc. ASRU}, 2017, pp. 712--718.

\bibitem{adiga2018use}
N.~Adiga, V.~Tsiaras, and Y.~Stylianou, ``On the use of {W}ave{N}et as a
  statistical vocoder,'' in \emph{Proc. ICASSP}, 2018, pp. 5674--5678.

\bibitem{ai2018samplernn}
Y.~Ai, H.-C. Wu, and Z.-H. Ling, ``{{SampleRNN}-based neural vocoder for
  statistical parametric speech synthesis},'' in \emph{Proc. ICASSP}.\hskip 1em
  plus 0.5em minus 0.4em\relax IEEE, 2018, pp. 5659--5663.

\bibitem{ai2019dnn}
Y.~Ai, J.-X. Zhang, L.~Chen, and Z.-H. Ling, ``{DNN}-based spectral enhancement
  for neural waveform generators with low-bit quantization,'' in \emph{Proc.
  ICASSP}, 2019, pp. 7025--7029.

\bibitem{lorenzo2018robust}
J.~Lorenzo-Trueba, T.~Drugman, J.~Latorre, T.~Merritt, B.~Putrycz, and
  R.~Barra-Chicote, ``{Robust universal neural vocoding},'' \emph{arXiv
  preprint arXiv:1811.06292}, 2018.

\bibitem{liu2018wavenet}
L.-J. Liu, Z.-H. Ling, Y.~Jiang, M.~Zhou, and L.-R. Dai, ``Wave{N}et vocoder
  with limited training data for voice conversion,'' in \emph{Proc.
  Interspeech}, 2018, pp. 1983--1987.

\bibitem{kobayashi2017statistical}
K.~Kobayashi, T.~Hayashi, A.~Tamamori, and T.~Toda, ``Statistical voice
  conversion with {W}ave{N}et-based waveform generation,'' in \emph{Proc.
  Interspeech}, 2017, pp. 1138--1142.

\bibitem{ling2018waveform}
Z.-H. Ling, Y.~Ai, Y.~Gu, and L.-R. Dai, ``{Waveform Modeling and Generation
  Using Hierarchical Recurrent Neural Networks for Speech Bandwidth
  Extension},'' \emph{IEEE/ACM Transactions on Audio, Speech, and Language
  Processing}, vol.~26, no.~5, pp. 883--894, 2018.

\bibitem{cui2018new}
Y.~Cui, X.~Wang, L.~He, and F.~K. Soong, ``A new glottal neural vocoder for
  speech synthesis,'' in \emph{Proc. Interspeech}, 2018, pp. 2017--2021.

\bibitem{juvela2018speaker}
L.~Juvela, V.~Tsiaras, B.~Bollepalli, M.~Airaksinen, J.~Yamagishi, and P.~Alku,
  ``{Speaker-independent raw waveform model for glottal excitation},'' in
  \emph{Proc. Interspeech 2018}, 2018, pp. 2012--2016.

\bibitem{valin2019lpcnet}
J.-M. Valin and J.~Skoglund, ``L{PCN}et: Improving neural speech synthesis
  through linear prediction,'' in \emph{Proc. ICASSP}, 2019, pp. 5891--5895.

\bibitem{wangNSFall}
\BIBentryALTinterwordspacing
X.~Wang, S.~Takaki, and J.~Yamagishi, ``{Neural Source-Filter Waveform Models
  for Statistical Parametric Speech Synthesis},'' \emph{IEEE/ACM Transactions
  on Audio, Speech, and Language Processing}, vol.~28, pp. 402--415, 2020.
  [Online]. Available: \url{https://ieeexplore.ieee.org/document/8915761/}
\BIBentrySTDinterwordspacing

\bibitem{ai2020neural}
Y.~Ai and Z.-H. Ling, ``{A neural vocoder with hierarchical generation of
  amplitude and phase spectra for statistical parametric speech synthesis},''
  \emph{IEEE/ACM Transactions on Audio, Speech, and Language Processing},
  vol.~28, pp. 839--851, 2020.

\bibitem{ai2020knowledge}
Y.~Ai and Z.-H. Ling, ``Knowledge-and-data-driven amplitude spectrum prediction
  for hierarchical neural vocoders,'' \emph{arXiv preprint arXiv:2004.07832},
  2020.

\bibitem{gunnar1960acoustic}
F.~Gunnar, \emph{The acoustic theory of speech production}.\hskip 1em plus
  0.5em minus 0.4em\relax The Hague, The Netherlands: Mouton, 1960.

\bibitem{zhao2019transferring}
Y.~Zhao, X.~Wang, L.~Juvela, and J.~Yamagishi, ``Transferring neural speech
  waveform synthesizers to musical instrument sounds generation,'' in
  \emph{Proc. ICASSP}, 2020, pp. 6269--6273.

\bibitem{engel2020ddsp}
J.~Engel, L.~Hantrakul, C.~Gu, and A.~Roberts, ``{DDSP: Differentiable Digital
  Signal Processing},'' \emph{Proc. ICLR}, 2020.

\bibitem{gulrajani2017improved}
I.~Gulrajani, F.~Ahmed, M.~Arjovsky, V.~Dumoulin, and A.~C. Courville,
  ``Improved training of {W}asserstein {GAN}s,'' in \emph{Advances in neural
  information processing systems}, 2017, pp. 5767--5777.

\bibitem{wang2019neural2}
X.~Wang, S.~Takaki, and J.~Yamagishi, ``Neural source-filter waveform models
  for statistical parametric speech synthesis,'' \emph{IEEE/ACM Transactions on
  Audio, Speech, and Language Processing}, vol.~28, pp. 402--415, 2019.

\bibitem{naylor2010speech}
P.~A. Naylor and N.~D. Gaubitch, \emph{{Speech dereverberation}}.\hskip 1em
  plus 0.5em minus 0.4em\relax Springer Science \& Business Media, 2010.

\bibitem{benesty2007springer}
J.~Benesty, M.~M. Sondhi, and Y.~Huang, \emph{{Springer handbook of speech
  processing}}.\hskip 1em plus 0.5em minus 0.4em\relax Springer, 2007.

\bibitem{valentini2018speech}
C.~Valentini-Botinhao and J.~Yamagishi, ``{Speech enhancement of noisy and
  reverberant speech for text-to-speech},'' \emph{IEEE/ACM Transactions on
  Audio, Speech, and Language Processing}, vol.~26, no.~8, pp. 1420--1433,
  2018.

\bibitem{kasi2002yet}
K.~Kasi and S.~A. Zahorian, ``Yet another algorithm for pitch tracking,'' in
  \emph{Proc. ICASSP}, vol.~1, 2002, pp. 361--364.

\bibitem{wang2020using}
X.~Wang and J.~Yamagishi, ``Using cyclic noise as the source signal for neural
  source-filter-based speech waveform model,'' \emph{arXiv preprint
  arXiv:2004.02191}, 2020.

\bibitem{gaubitch2012performance}
N.~D. Gaubitch, H.~W. Loellmann, M.~Jeub, T.~H. Falk, P.~A. Naylor, P.~Vary,
  and M.~Brookes, ``Performance comparison of algorithms for blind
  reverberation time estimation from speech,'' in \emph{Proc. IWAENC}, 2012,
  pp. 1--4.

\bibitem{marco2020blind}
\BIBentryALTinterwordspacing
M.~Jeub, ``Blind reverberation time estimation,'' 2015. [Online]. Available:
  \url{https://www.mathworks.com/matlabcentral/fileexchange/35740-blind-reverberation-time-estimation}
\BIBentrySTDinterwordspacing

\bibitem{buchholz2011crowdsourcing}
S.~Buchholz and J.~Latorre, ``Crowdsourcing preference tests, and how to detect
  cheating,'' in \emph{Proc. Interspeech}, 2011, pp. 3053--3056.

\bibitem{recommendation2001method}
I.~Recommendation, ``Method for the subjective assessment of intermediate sound
  quality ({MUSHRA}),'' \emph{ITU, BS}, pp. 1543--1, 2001.

\end{thebibliography}

\end{document}